\documentstyle[11pt,newpasp,twoside,psfig]{article}

\newcommand{\ha}{H\,$\alpha$}
\newcommand{\hb}{H\,$\beta$}

\newcommand{\civ}{C~{\sc iv}}

\newcommand{\lya}{Ly\,$\alpha$}

\newcommand{\ew}{$W_{\lambda}$}

\newcommand{\aopt}{$\alpha_{\rm opt}$}

\newcommand{\zaze}{$z_{\rm a} \approx z_{\rm e}$}

\newcommand{\kms}{km\,s$^{-1}$}

\begin{document}

\title{Associated absorption and radio source growth}
\author{ Joanne C. Baker}
\affil{Astronomy Department, 601 Campbell Hall, 
University of California, Berkeley CA 94720, USA}

\begin{abstract}

Results are presented from a survey for \civ\ associated absorption 
in a complete low-frequency-selected  sample of quasars. 
In agreement with previous work, associated absorbers are 
most common in steep-spectrum and lobe-dominated quasars,
indicative of an anisotropic cloud distribution. 
Furthermore, we find the strongest \civ\ absorption 
occurs in  sources of small radio size, suggesting that 
the absorbing clouds are destroyed or
displaced as radio sources expand. Evidence for dust in the
clouds is also found, such that quasars with 
strong absorption are systematically redder. Finally, we find
no evidence for evolution in the frequency or properties of
the absorbers from $z\sim0.7$ to $z\sim3$. 

\end{abstract}


\section{Introduction}

Absorption lines occurring very close to the 
quasar redshift --- associated absorbers --- are potentially 
valuable probes of quasar environments. The precise origin of the
absorbing material is unknown, but many interesting 
regions along the sightline may contribute. Absorption may arise, 
for instance, in gas near the quasar nucleus, in the ISM of 
the host galaxy or in neighbouring galaxies.

It seems clear that the majority of associated absorption systems 
(as defined by $|z_a - z_e|<5000$\,\kms) are 
related directly to the quasar phenomenon, rather than being
due solely to cosmologically-distributed foreground galaxies.
First, the density of associated absorption systems 
per redshift interval is greater than expected for 
intervening galactic systems alone 
(Foltz et al. 1988, Richards et al. 1999, 2001).
Secondly, the characteristics of the absorbers depend on quasar
type. Narrow absorption lines occur more frequently in 
radio-loud quasars, especially those with steep radio spectra 
(Anderson et al. 1987; Foltz et al. 1988; Richards et al. 1999, 2001).
Alternatively, optically-selected samples might be biased against 
objects with absorption. Previous studies have interpreted the 
prevalence of \zaze\ systems
in steep-spectrum and lobe-dominated quasars in terms of orientation 
(Barthel, Tytler \& Vestergaard 1997). In addition, \zaze\ systems 
differ subtly in ionisation or velocity  profiles from those 
seen along sightlines traversing normal galaxy halos, consistent
with their proximity to the AGN (Hamann \& Ferland 1999). 

A major limitation of most previous studies of \zaze\ absorption 
is that they have used inhomogeneous samples which are prone to
strong selection effects. To counter this, we
have observed \zaze\ absorption in quasars drawn from a complete sample of 
408-MHz selected quasars. Low-frequency radio selection and a
high completeness level ensure that orientation bias is minimised and
reddened sightlines are included. The initial results 
of this study, correlations between \civ\ absorption and quasar 
properties, are summarised here. Full results will be published shortly
(Baker et al. 2001, in preparation).
Cosmological parameters $H_0=50$ \kms Mpc$^{-1}$, $\Omega=1.0$ and 
$\Lambda = 0$ are assumed for consistency with our earlier work.

\section{The quasar sample and observations}

We have obtained intermediate-resolution spectroscopy of 
the \civ\ to \lya\ spectral region for 43 quasars drawn from the 
complete Molonglo Quasar Sample (MQS;
Kapahi et al. 1998; Baker et al. 1999).   
Briefly, the MQS comprises all quasars 
in the $-30^{\circ}< \delta < -20^{\circ}$ strip of the 408-MHz 
Molonglo Reference Catalogue (MRC; Large et al. 1981) down to a 
limiting flux density of $S_{408}=0.95$~Jy. The MQS 
contains 111 quasars  with $0.1<z<3.0$. 

MQS quasars were selected for the absorption study in two redshift ranges, 
$1.5<z<3.0$ where redshifted \civ\ is observable from the ground, 
and $0.7<z<1.0$ which was observed in the UV using STIS on HST.
Ground-based spectra with spectral resolution 1--2.4\AA\ (FWHM)
were obtained for  22 out of  a total of 
27 MQS quasars with $z>1.5$, using mostly  the 
Anglo-Australian Telescope (AAT) and the ESO 3.6m telescope. 
In addition, four faint quasars 
were observed with FORS1 on the VLT (UT1). 
The STIS spectroscopy was carried out between May 1999 and February 2001
for 19 MQS quasars with redshifts $0.7<z<1.0$. The NUV-MAMA detector was 
used with the G230L grism, 
giving a spectral resolution of 3.0\AA\ over the wavelength 
range 1570--3180\AA.

Absorption systems were identified and measured using IRAF. 
The strongest absorption system was identified in each spectrum
within $\pm5000$\kms\ of the \civ\ emission-line redshift. 
The majority  (50--70\%) of the resulting systems lay
within $\pm500$\kms\ of the emission redshift, and were
both blue- and red-shifted. 

\section{Results}

\subsection{Radio spectrum and morphology}

Although not shown explicitly in this short contribution, 
we do  confirm the trends for absorption to be most prevalent in 
steep-spectrum and lobe-dominated quasars
(e.g. Anderson et al. 1997; Foltz et al. 1988; 
Barthel et al. 1997). In our MQS study,  for example, 
strong absorption (\ew$>1$\AA) was detected 
exclusively in steep-spectrum ($\alpha > 0.5$) quasars in both the 
high- and low-redshift sub-samples. 

\begin{figure}
\centerline{\psfig{file=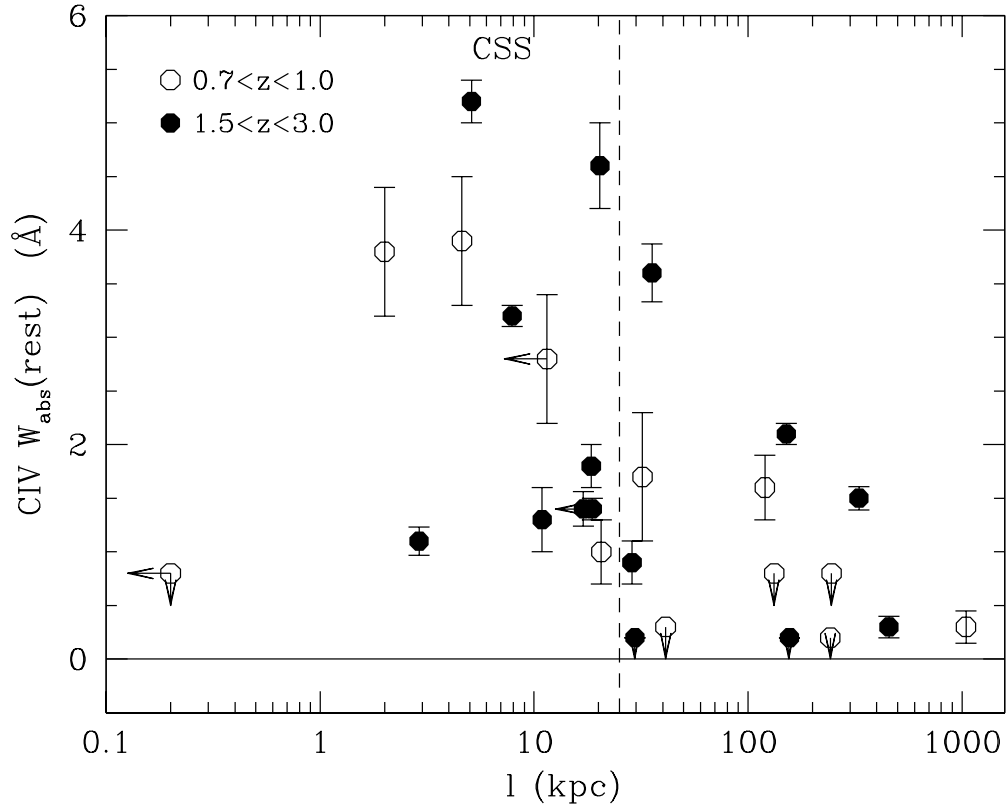,bbllx=
20pt,bblly=140pt,bburx=575pt,bbury=705pt,height=14cm} }
\caption[e]{
Equivalent width of \civ\ absorption as a function of 
radio source size, $l$ (kpc).  
Quasars with $0.7<z<1.0$ are plotted as open circles, 
those with $1.5<z<3.0$ are plotted with filled symbols.
The dotted line at $l=25$ kpc
illustrates our working definition of CSSs. 
Arrows indicate limits. 
}
\label{ewl}
\end{figure}

\subsection{Radio size}

In the MQS data, the most striking result is that the strongest 
absorption occurs preferentially in the smallest radio sources.
The equivalent widths of \civ\ absorption  are plotted 
as a function of radio source size in Figure \ref{ewl} 
for both high- and low-redshift datasets. Highly beamed 
core-dominated quasars are excluded as they are expected to be 
severely foreshortened. Compact, steep-spectrum (CSS) sources
are included on the plot.  
CSS sources (see review by O'Dea 1998) 
are intrinsically small ($l<25$ kpc) with
steep radio spectra ($\alpha>0.5$ where $S_{\nu} \propto \nu^{-\alpha}$).
The precise definition of CSSs is somewhat arbitrary, 
comprising essentially those sources whose radio
emission was unbeamed (not core-dominated) yet
unresolved with conventional arrays (arcsec resolution). 
Thus CSS sources do include intrinsically small, and perhaps young, 
sources. 
Radio sizes for the CSS quasars in the MQS study were measured from MERLIN
images with $\sim 0.1''$  resolution (de Silva et al. 2001, in preparation). 
Notably, all the CSS quasars in our study (except an
unusual GigaHertz-Peaked Spectrum source) show \zaze\ absorption
stronger than $W_{\rm abs}=1$\AA.

\begin{figure}
\centerline{\psfig{file=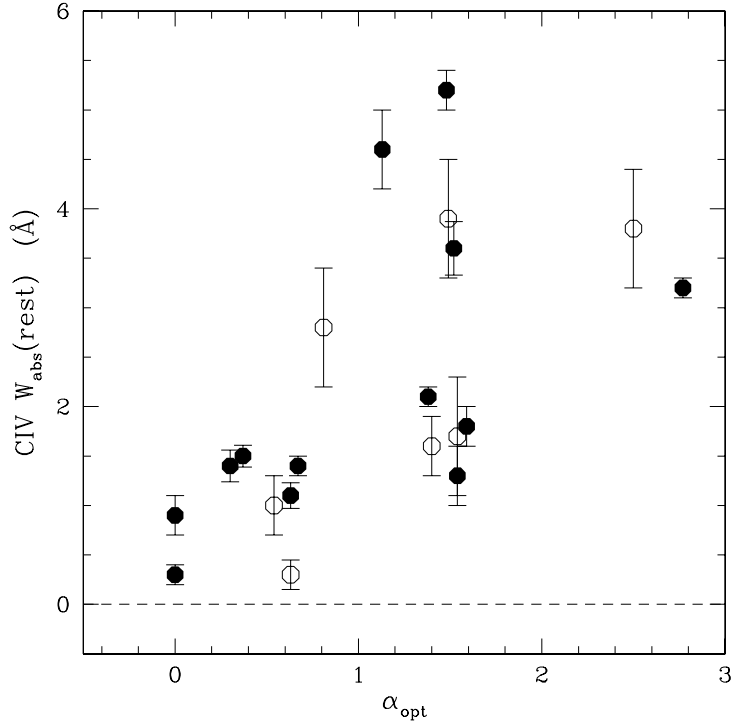,bbllx=
20pt,bblly=140pt,bburx=575pt,bbury=705pt,height=10cm} }
\caption[e]{
Equivalent width of \civ\ absorption as a function of 
optical spectral index, for all quasars with detected absorption. 
Quasars with $0.7<z<1.0$ are plotted as open circles, 
those with $1.5<z<3.0$ are plotted with filled symbols.
}
\label{ewaopt}
\end{figure}

\subsection{Reddening by absorbing clouds}

For quasars where  \civ\ absorption was detected, the equivalent
width of the absorption is plotted in Figure \ref{ewaopt} against 
the slope of the optical spectrum, $\alpha_{\rm opt}$ (as 
observed between 3500 and 10000\AA). There is a strong 
correlation between absorption-line strength and spectral 
slope --- heavily absorbed quasars are systematically redder.

Baker \& Hunstead (1995) and 
Baker (1997) presented evidence that the range in optical spectral slope 
observed in the MQS is due in part to reddening by an anisotropic dust 
screen lying outside the broad emission-line region. 
In this earlier study, the most direct evidence for dust reddening 
(as opposed to intrinsic spectral steepening) was the tight correlation 
between \aopt\ and broad \ha/\hb\ Balmer Decrement,  
at least in low-redshift quasars where it was measurable in the optical.
The reader is referred to Baker (1997) for this result and 
a more detailed description of the dust-reddening hypothesis. 
By extension, the simplest explanation of the 
correlation of $W_{\rm abs}$ with \aopt\ is that
the absorbing gas clouds contain dust, and they lie outside the 
nuclear continuum source.
Alternatively, if dust is not responsible for the red continuum slopes, 
then \civ\ absorption strength correlates with an intrinsically 
softer continuum shape.

\section{Discussion}

These results suggest that the distribution of associated absorbers 
in quasars is dependent on {\it both orientation and radio size\/}. 
Orientation explains the trends with
radio spectral index and radio-core dominance, as described by 
Barthel et al. (1997). Orientation, however, cannot explain the 
stronger absorption in CSS sources and the global decrease 
in absorbing column density with radio size. 

The radio-size dependence of associated absorption may be explained
if the absorption column density either correlates with quasar 
environmental density, or changes with time.
Currently, the first hypothesis is not supported by observations, which find 
that CSS host galaxies appear to be the same as those harbouring 
larger sources, and CSS quasars do not systematically reside 
in clusters more often than larger sources (de Vries et al. 2000;
O'Dea 1998 and references therein). 
The alternative idea is that the absorbing column density
decreases as the radio source grows. This could occur 
because of photoionisation of the clouds by the quasar over time, or
by direct interaction of the radio jet and its cocoon 
on the absorbing clouds, or both. 

The strong correlation between the \civ\ absorption strength and 
red continuum in the quasars is highly suggestive of dust 
in the absorbing clouds. However, a cospatial distribution
of dust and gas is problematic, dust should be destroyed by
sputtering in the hot gas where \civ\ absorption arises.
De Young (1998) points out that the strong shocks in radio-source
environments should destroy dust easily on timescales $\ll 10^{6}$ yrs,
which is much shorter than the lifetime of the radio source. 

Putting all the evidence together, we propose a consistent model 
whereby radio sources are born enshrouded in dust and gas, which is
gradually destroyed and ionised (respectively)  along the
radio axis as the source expands. 

In addition, we find no evidence for changes in the the frequency
or strength of the absorbers with redshift from $z\sim 0.7$ to $z\sim 3$.
This lack of evolution is perhaps unexpected given the absorbers 
are probably at kpc distances where they should be affected by
quasar environmental and perhaps galactic evolution.

\section{Conclusions}

Initial results from a study of \civ\ associated absorption in a
complete, homogeneous sample of radio-loud quasars are presented.
The results confirm that the absorbing cloud distribution is 
anisotropic, such that absorption is more common in 
steep-spectrum and lobe-dominated quasars. Furthermore, 
we find new evidence that the strength of \civ\ absorption decreases
with increasing radio source size. If we assume that the larger 
sources are older than the smaller (CSS) ones, then we can
attribute the decrease in column density to the
growth of the radio source envelope through the ISM.
The absorbing clouds probably contain dust, which reddens the quasar light.
Consequently we predict that absorbed
quasars will be missed preferentially in optically-selected samples. 
Finally, these results appear to be independent of redshift,
giving essentially the same picture at $z\approx 0.7$ and $z\sim 3$, 
epochs between which evolution of quasar environments should be discernible. 
Thus we are drawn to a picture where radio sources are born in
gaseous and dusty cocoons, from which they emerge as their
radio jets expand beyond the host galaxy.

\acknowledgments

JCB acknowledges support by NASA through Hubble Fellowship grant 
\#HF-01103.01-98A from STScI, 
which is operated by the AURA, Inc., under NASA contract NAS5-26555.

\end{document}